\documentstyle[prl,aps,multicol]{revtex}

\newcommand{\ba}{\begin{array}}
\newcommand{\ea}{\end{array}}

\newcommand{\be}{\begin{equation}}
\newcommand{\ee}{\end{equation}}
\newcommand{\bea}{\begin{eqnarray}}
\newcommand{\eea}{\end{eqnarray}}
\newcommand{\beal}{\setcounter{letter}{1} \begin{eqnarray}}
\newcommand{\eeal}{\addtocounter{equation}{1} \end{eqnarray}}

\newcommand{\req}[1]{Eq.(\ref{#1})}

\newcommand{\larrow}{\,\,\,\,\hbox to 30pt{\rightarrowfill}
\,\,\,\,}
\newcommand{\slarrow}{\,\,\,\hbox to 20pt{\rightarrowfill}
\,\,\,}

\def\be {\begin{equation}}
\def\ee {\end{equation}}
\def\ba {\begin{eqnarray}}
\def\ea {\end{eqnarray}}

\def\c  {\gamma}
\def\d  {\delta}

\def\O  {\Omega}

\def\le {\left}
\def\ri {\right}
\def\pa {\partial}
 
\begin{document}
 
\title{Scalar field spacetimes and the AdS/CFT conjecture}
\author{Saurya Das $^\dagger$, 
J. Gegenberg $^\sharp$, and 
V. Husain $^\sharp$} 
\address{$\dagger$  
Department of Physics, The University of Winnipeg\\
515 Portage Avenue, Winnipeg, Manitoba, CANADA  R3B 2E9\\
$\sharp$ Dept. of Mathematics and Statistics,
University of New Brunswick\\
Fredericton, New Brunswick, CANADA  E3B 5A3\\
emails: saurya@theory.uwinnipeg.ca, lenin@math.unb.ca,husain@math.unb.ca}

\date{\today}

\maketitle

\begin{abstract}
{We describe a class of asymptotically AdS scalar field spacetimes, and 
calculate the associated conserved charges for three, four and five 
spacetime dimensions using the conformal and counter-term prescriptions. 
The energy associated with the solutions in each case is proportional to 
$\sqrt{M^2 - k^2}$, where $M$ is a constant and $k$ is a scalar charge. 
In five spacetime dimensions, the counter-term prescription gives an 
additional vacuum (Casimir) energy, which agrees with that found in the 
context of AdS/CFT correspondence. We find a surprising degeneracy: 
the energy of the ``extremal'' scalar field solution $M=k$ equals the 
energy of global AdS. This result is discussed in light of the AdS/CFT 
conjecture.}

\end{abstract}

\begin{multicols}{2}

The non-linear coupling of gravity to matter in general relativity
presents difficult technical problems in attempts to understand 
gravitational interactions of elementary particles and strings, as well as 
questions such as the details of gravitational collapse. Progress in the 
former area has come  mainly from treating quantum fields as propagating 
on fixed background geometries \cite{gswp}, whereas much of the progress 
in the latter has come from detailed numerical work 
\cite{chop}. 

Exact solutions of the relevant matter-gravity equations can play an important 
role by shedding light on questions of interest in both general relativity and 
string theory. One is often interested in certain classes of solutions, with 
specified asymptotic properties, the most common of which are the asymptotically 
flat spacetimes. Recent work in string theory however has highlighted  
the importance of another class of spacetimes via the AdS/CFT conjecture 
\cite{maldarev}. These are the asymptotically anti-deSitter spacetimes (AAdS). 

The AdS/CFT conjecture is a duality between string theory 
on $AdS_5\times S^5$ and the large $N$ limit of conformally invariant 
${\cal N}=4$ 
$SU(N)$ Yang-Mills (YM) theory on the boundary of $AdS_5$. This conjecture
proposes a direct correspondence between physical effects associated
with fields propagating in AdS spacetime and those of a conformal
quantum field theory on the boundary of AdS spacetime.  Significant
evidence for the conjecture 
has come from studying free scalar or other fields on a fixed $AdS$ 
background \cite{adsmatter}. 
Another important aspect of this conjecture is that it connects infrared 
effects in AdS space to ultraviolet effects in the boundary theory. This in 
turn implies a 
connection between low and high energy physics in the respective theories. 
A fully non-linear gravity-scalar field solution can provide a window into 
this aspect of the conjecture as well, as is the case for scalar fields on 
a fixed background.  

With this motivation we study the Einstein-scalar field system, with
minimally coupled massless scalar field and negative cosmological constant. 
We present static spherically symmetric  AAdS solutions of these equations. 
For spacetime dimension $d=3$, the equations can be solved exactly. 
For $ d \geq 4$,  the corresponding equations can be solved analytically 
for large radial distance, ie. asymptotically. We calculate conserved charges 
associated to these spacetimes using the conformal \cite{am} and the 
counterterm \cite{bk1} prescriptions. Finally we discuss some 
consequences of our results, in particular the surprising energy 
degeneracies associated with the solutions: The ``extremal'' limit of 
our solutions have the same energy as the corresponding global $AdS_d$ 
spacetime.

The solutions we discuss are singular at the origin. This raises the 
 question of whether these are ``admissable'' in the context of the AdS/CFT 
conjecture. The fact that the energy of this class of solutions turns out 
to be finite is a hint that the singularity may be resolved by quantum 
effects. Indeed, there has been a suggestion \cite{hm} that singularities 
are valuable in that they are concomitant with the absence of an energy 
lower bound, which is another criteria for excluding solutions.
This is not the case for the solutions we present, since the corresponding 
energies are well defined and bounded below by the energy of global AdS.

The equality of the vacuum energy of ADS (calculated via the counter-term 
method), and the Casimir energy of the boundary YM theory is considered to be 
a  piece of the evidence for the AdS/CFT conjecture. However, our degeneracy 
result demonstrates that {\it there is another spacetime with the same 
Yang-Mills Casimir energy}. This raises an ambiguity concerning this 
dictionary entry of the AdS/CFT correspondence. 
 
The $d=3$ matter-gravity model considered in this paper 
has been studied numerically in the full time dependent context with 
circular symmetry in Refs.  \cite{pc,vh}, where critical behavior 
at the onset of black hole formation is observed. More recently the 
critical exponent for the apparent horizon radius found in \cite{vh} has 
been verified by an analytical perturbation theory calculation \cite{cf}.

We consider the massless scalar field minimally coupled to gravity in
$d$ spacetime dimensions. With the parametrization of the
negative cosmological constant as
$\Lambda = - (d-1)(d-2)/2\ell^2$, the field equations are
 \be
R_{ab} + {(d-1)\over \ell^2}\ g_{ab} =   \pa_a\phi \pa_b\phi.
\ee
where factors of $8\pi G$ are suppressed. We assume the static spherically 
symmetric form  
\be
ds^2  = -g(r)h(r) dt^2 + {h(r)\over g(r)}dr^2
+ r^2 d\Omega_{d-2}^2,
\label{metric}
\ee
of the metric in $d$ spacetime dimensions, which give the coupled 
equations 
\ba
r \ell^2 g' &+& (d-3) (g-h) \ell^2 - (d-1) r^2 h = 0, \nonumber \\
(\ln h)' &=& \frac{r}{d-2} (\phi')^2, \label{diff1}\\
\phi ' &=& \frac{k}{g r^{d-2}},  \nonumber
\ea
where the prime $'$ denotes the derivative with respect to $r$ and $k$ is an
integration constant obtained by integrating the Klein-Gordon equation.

By a field redefinition $ g(r) := \psi(r)/r^{d-3}$, the above set of
equations give 
\bea
&& \psi^2\le[ r \frac{\psi''}{\psi'}  -
\frac{(d-1)(d-2) r^2 + (d-3)(d-4) \ell^2}{ (d-1)r^2 + (d-3)\ell^2}
\ri] \nonumber \\
&& = \frac{k^2}{d-2}
\label{master}
\eea
This basic equation determines the spacetime
metrics of interest. This equation is especially simple for $d=3$ and
we deal with it separately, followed by the cases $d \geq 4$.

$d=3$: The differential equation (\ref{master}) reduces to 
\cite{burko}:
\be
g(r)^2 \left[ r \frac{g''(r)}{g(r)}  - 1  \right] = k^2
.\ee
The complete solution is this case can be expressed as $r=r(g)$ with
\be
r = B~\ell \left(g^2 + Ag/2 - k^2/2 \right)^{1/4}
\left( g+A/4+C/4\over g+A/4-C/4\right)^{A/4C}
\label{gensol}
\ee
where $A$ and $B>0$ are integration constants, and
$C=\sqrt{8k^2 + A^2}$. Without loss of generality, we can set $B=1$.
 It is convenient to define the variables $x=g+A/4, b=C/4,a=A/C$,
in terms of which \req{gensol} becomes
\be
r=\ell~(x-b)^{(1-a)/4}(x+b)^{(1+a)/4} .
\ee
Using (\ref{diff1}) gives
\ba
h(x) &=& \frac{x^2-b^2}{x-ab} (x-b)^{-(1-a)/2} (x+b)^{-(1+a)/2}
,\nonumber \\
\phi(x) &=& \frac{1}{2}{\sqrt{\frac{1-a^2}{2}}}
\ln \le(\frac{x-b}{x+b}\ri). \nonumber
\ea
Thus the scalar field is real for $a^2 \leq 1$. The metric
may be written in the new radial coordinate $x$ as
\ba
ds^2 &=& - (x-b)^{(1+a)/2}(x+b)^{(1-a)/2} dt^2 +
 {l^2\over 4(x^2-b^2)} dx^2  \nonumber \\
       &+& \ell^2~(x-b)^{(1-a)/2}(x+b)^{(1+a)/2} d\O_{d-2}^2.
\label{newsol}
\ea
The Ricci scalar of the metric is
\be
R  = -2\frac{3x^2-4b^2+a^2b^2}{\ell^2(x^2-b^2)}
\ee
which shows that there are curvature singularities at $ x= \pm b$,
corresponding to the origin $r=0$. Also, it confirms that the
spacetime is AAdS, since $R (x \rightarrow \infty) = - 6/\ell^2$.
Since the solution contains no horizons for  non-vanishing scalar 
field, the singularity at $r=0$ is naked.

There are two special cases of this metric which are
familiar, both of which correspond to vanishing scalar
field, $k=0$. The first is $a^2=1$ for which the metric
reduces to
\be
ds^2 ~=~ - \left(\frac{r^2}{\ell^2} \mp 2b\right) dt^2 +
\frac{dr^2}{\frac{r^2}{\ell^2} \mp 2b} + r^2 d\theta^2,
\ee
where the $\mp$ signs correspond to $a= \pm 1$ respectively.
Thus $a=1$ is the non-rotating BTZ black hole with mass $2b=C/2$.
The second is $b=0$ and $a$ arbitrary, which gives the zero mass
BTZ black hole, rather than global $AdS_3$ spacetime.

$d \geq 4$: For spacetime dimensions greater than three, 
equation (\ref{master}) cannot be solved analytically. However it is possible 
to obtain an asymptotic expansion of the solution for large $r$.  For
$r \gg \ell$ (and fixed $\ell$), (\ref{master}) can be approximated as:
\be
\psi^2 \le[ \frac{\psi''}{\psi'} r - (d-2) \ri] = \frac{k^2}{d-2}
,\label{master1}
\ee
which has the exact solution $\psi_0(r) $ given implicitly by
\ba
r &&= B~\ell^{2/(d-1)}
 \left[ \psi_0^2 + \frac{A\psi_0}{d-1} - \frac{k^2}{(d-1)(d-2)}
\right]^{\frac{1}{2(d-1)}} \times \nonumber \\
 &&\left[
\psi_0 + A/2(d-1) + C/2(d-1)(d-2) \over
\psi_0 + A/2(d-1) - C/2(d-1)(d-2)
\right]^{\frac{A}{2C}\left(\frac{d-2}{d-1}\right)}
\label{gensol2}
\ea
As before the metric for large $r$ may be written 
using the variables $x = \psi_0 +  A/2(d-1)$,  
$b =  C/2(d-1)(d-2)$,  and $a = (d-2) A/C$.  
For large $r$, $\psi_0(r)\sim r^{(d-1)}$.

The next term in the asymptotic expansion for large $r$ is
obtained by writing
\be
\psi(r)=\psi_0(r)(1+\alpha \ell^2/r^2),
\ee
where $\alpha$ is a constant. For $d=5$ one finds $\alpha =1$
by substituting this expression into \req{master}, a result which
is useful for calculating conserved charges.

For AAdS geometries, the 
calculation of conserved charges is complicated by the occurrence 
of divergent expressions. These occur essentially because the metric 
diverges as $r^2$ for large $r$. There exist two quite distinct procedures 
for obtaining ``regularized'' finite  expressions for asymptotic conserved 
charges. These are the so-called ``conformal'' \cite{am} and ``counter-term'' 
\cite{bk1} methods.  

In the conformal method, a conformal transformation 
$g_{ab} = \Omega^2 \hat{g}_{ab}$ is performed on
the physical spacetime $(\hat{M}, \hat{g}_{ab})$ under consideration, 
such that the asymptotic regions get mapped to a finite distance in a 
new manifold $(M,g_{ab})$ \cite{am}. A boundary  is then
added to this conformally transformed (and unphysical) manifold.
This is especially useful for AAdS spacetimes, because many of the 
canonical metric components diverge asymptotically and
limits such as $r \rightarrow \infty$ become rather tricky. On the other
hand, the transformed manifold has a completely well-behaved structure.

The procedure basically involves showing that the electric parts
of the Weyl tensor satisfy, as a consequence of Einstein's 
equations, a conservation law at null infinity, $\mathcal{I}$, in the 
conformally transformed spacetime. The end result is that the following 
equation holds on $\mathcal{I}$:
\begin{equation}
D^{p} E_{mp}=\,-{8\pi G}\,\,\frac{(d-3)}{\ell }\,\, T_{ab}n^{a}
h^{b}{}_{m}\,.
\end{equation}
where $D^{p}$ is the intrinsic covariant derivative on $\mathcal{I}$,
compatible with the induced metric $h_{ab} := g_{ab}-{\ell}%
^{2}n_{a}n_{b}~~(n_{a}:=\nabla _{a}\Omega )$, $ E_{ab}$ is the
electric part of the Weyl tensor at $\mathcal{I}$ defined as:
$ E_{ab}:={\ell}^{2}~\Omega ^{3-d}C_{ambn}n^{m}n^{n}$
and $ T_{ab}:=\Omega^{2-d}{\hat{T}}_{ab}$ on $\mathcal{I}$.
The above conservation equations dictate the following form of
conserved charge associated with the conformal killing vector field (KVF)
$\xi $:
\begin{equation}
Q_{\xi }[C]:=-\frac{1}{8\pi G}~\frac{\ell
}{d-3}~~\oint_{C} E_{ab}\xi^{a}dS^{b}.  \label{conserve}
\end{equation}
(An ordinary KVF on ${\hat M}$ becomes the conformal KVF on
$M$.) In the presence of matter fields, this charge satisfies the  
covariant balance equation
\begin{equation}
Q_{\xi }[C_{2}]-Q_{\xi }[C_{1}]=\int_{\Delta
\mathcal{I}} T_{ab}\xi^{a}dS^{b}  
\label{balance}
\end{equation}
where $C_{1}$ and $C_{2}$ are two cross-sections on
$\mathcal{I}$ bounding the region $\Delta \mathcal{I}$. 
Equations (\ref{conserve}) and (\ref{balance}) are the fundamental 
relations which we will use to define conserved quantities.
Thus apart from volume factors, the electric part of the Weyl tensor
is the relevant quantity to be calculated. For our metrics the scalar 
field vanishes too quickly to be captured on the right hand side 
of Eqn. (\ref{balance}). 

The counter-term method proposes that the Einstein-Hilbert action  
$S_{EH}$ should be supplemented with additional boundary terms dependent on 
the intrinsic  metric. Since the variational principle is
defined with fixed boundary metric, this does not change the equations
of motion.  The full action  $S + S_{ct}$ is used to obatin an effective 
energy momentum tensor associated with the boundary:
\be
T_{ab} := \frac{2}{\sqrt{ -\c}} \frac{\d (S_{EH}+S_{ct})}{\d \c^{ab}}
.\label{tab}
\ee
The conserved charge associated with the symmetry generated by 
a vector field $\xi$ is then defined by 
\be
Q_\xi := \frac{1}{8\pi G}\int_\Sigma d^{d-2} {{\sqrt \sigma}}~T_{ab} u^a
\xi^b,
\label{countercharge}
\ee
where $\Sigma$ is a spatial slice of $\partial M$, $u^a$ is the timelike 
unit normal at $\Sigma$, and $\sigma_{ab} := \c_{ab} + u_au_b$. 

For $d=3,4$ and 5, the following expression suffices to yield finite charges 
for commonly encountered AAdS spacetimes, including the scalar solutions:
\be
 {L}_{ct} = -\frac{d-2}{\ell }\sqrt{-\gamma }-\frac{\ell
\sqrt{-\gamma }}{2(d-3)} R(\gamma).\label{CTaction}
\ee
The resulting stress-energy tensor is
\bea
T_{ab}&=&K_{ab}-\gamma _{ab}K-\frac{d-2}{l}\gamma _{ab}\nonumber\\
       &&+\frac{l}{d-3}%
\left(  R_{ab}(\gamma)-\frac{1}{2}\gamma _{ab} R(\gamma)\right)
.\label{BYtensor}
\eea

The conserved charges for various dimensions are calculated
using (\ref{conserve}) in the conformal method ($Q_1$), and 
(\ref{countercharge}) in the counterterm method ($Q_2$). 
Restoring the $8\pi G$ factors we obtain the following 
expressions.
 
\begin{center}

\begin{tabular}{|c|c|c|}
\hline
Dim & $Q_1$& $Q_2$  \\
\hline
3 &  - & $ab/4 G $    \\
\hline
4 & $ab/G $ & $ab/G$     \\
\hline
5 &  $9 \pi ab/16G$  &
$9\pi ab/16G + 3\pi l^2/32G$  \\
\hline
\end{tabular}

\vspace{.2cm}
  Conserved quantities in the two approaches.
\end{center}
Note that the conformal method does not apply for $d=3$, as the Weyl
tensor is identically zero. The following comments discuss additional 
aspects of the solutions: 

\noindent (i) Although we have not been able to find an 
exact solution of (\ref{master}) above three dimensions for 
all $r$, there is strong evidence that the
solutions in higher dimensions have curvature 
singularities at the origin. This may be seen analytically 
and numerically. Analytically, (\ref{master}) can be 
integrated near $r=0$. For example, in $d=5$ this equation 
gives 
\be
\psi=\sqrt{{k^2\over 6}-{r^4\over 6 \ell^4}},
\ee 
from which the Ricci scalar is $R \sim r^{-6}$. Numerically, this 
equation can be integrated for all $r$ and $d$ (for example using MAPLE) 
\cite{maple}. For all initial conditions considered, $\phi'$ shows 
a strong divergence at the origin, which is similar to the behaviour 
seen analytically in three dimensions. Finally, an exact scalar field
solution in five dimensions has been found recently by one of us 
\cite{vcgr}. Although this exact solution arises in different 
coordinates, it has the same symmetries as the class considered in 
this paper. This provides further evidence that the spherically 
symmetric scalar field-AdS spacetimes generically have naked 
singularties at the origin. 

\noindent (ii) The two approaches for calculating conserved charges give
positive energies for our scalar field solutions in spite of the 
fact that the solutions contain a naked singularity at the origin. 
This is unlike the negative mass Schwarzschild naked singularity, 
where the associated conserved charge is negative. 
 
\noindent(iii) Except in $d=5$, the two approaches yield identical results.
In $d=5$, the counterterm prescription predicts an additional Casimir
energy, which is identical to that found in the context of AdS/CFT
correspondence, and is independent of the scalar charge $k$. 
This is unlike AAdS spacetimes with rotation, where the 
vacuum energy depends on the rotation parameter \cite{am}.
Our results can be generalized to spacetime dimensions greater than
five; the Casimir energy  appears for all odd
dimensional spacetimes. 

\noindent(iv) In terms of the scalar field strength $k$, the 
conserved charges are proportional to 
\be 
ab = {1\over 4}\sqrt{ {C^2\over (d-2)^2} - 4k^2\ {(d-1)\over (d-2)} },
\ee 
Note that the $k=0$ case is the AdS-Schwarzschild solution with 
mass $M = C/(d-2)$. Thus, surprisingly the presence of the scalar 
field effectively reduces the energy of the gravitational solution.    

\noindent(v) There is an unexpected and interesting result in 
the ``extremal'' case  
\be 
      {C\over d-2} = 2k \sqrt{{d-1\over d-2}}. 
\ee
The conserved charge vanishes in the conformal calculation for all dimensions
and is therefore equal to the energy of global AdS in this method. 
The same holds for the counter-term method, except that the energies of the 
two solutions now equal the Casimir energy in five dimensions. This surprising 
degeneracy of energy associated with two distinct solutions on the gravity 
side raises an interesting question for the AdS/CFT conjecture:   
What effects on the CFT side distinguish global AdS from this extremal 
scalar field solution? It is possible that the answer lies in calculating 
other effects using the correspondence, such as n-point functions with  
these spacetimes as background (see below). 

\noindent(vi) The energy of Schwarzschild-AdS spacetime $M_{SA}$ 
(calculated using either method) can be matched by choosing $C$ and $k$ of 
the scalar solutions such that 
\be
M_{SA} = {1\over 4}\sqrt{{C^2\over (d-2)^2} - 4k^2\ {(d-1)\over (d-2)} }.
\ee
Thus the energy degeneracy pointed out above, concerning global AdS,  
extends to the Schwarzschild-AdS metrics. 

\noindent(vii) There exist certain intuitive criteria 
for the types of gravitational singularities that might be permitted in 
the context of the AdS/CFT conjecture\cite{gub}. The central singularity 
in the present solutions does not violate these. Thus it appears 
that an explanation of the degeneracy is not yet available.  

\noindent(viii) Some evidence that the energy degeneracy discussed 
above may not be manifested in other calculations in the AdS/CFT context 
comes from consideration of the full action associated with our solutions.
This computation can be done exactly in three dimensions. 
Surprisingly, the divergence at the origin does not make the action
infinite. (Recall that the  divergence for large $r$ is regulated 
by the subtraction procedure). The key test is whether the scalar field 
parameter appears in the action. It does:  $S = -\pi k$ (multiplied 
by a factor coming from the $t$ integration) for the ``extremal'' 
solution. (For comparison the action of global AdS is $-2\pi$.) 
This situation is analagous to the result for  Schwarzschild-AdS, where 
the action is a function of the black hole mass $M$. 
The natural interpretation of the latter \cite{witbh,maldarev} is that 
it corresponds to a CFT at finite temperature. Since there
is a scale in our solution, determined by the scalar field strength,
the correponding CFT must have conformal invariance broken. The 
exact nature of the breaking is apparently not due to a temperature
since temperature cannot be associated with naked singularities.  
Nevertheless, these considerations provide a clear distinction between 
global AdS and this class of scalar field spacetimes, regardless of the 
energy degeneracy. 

In summary we have described solutions of general relativity
with a cosmological constant coupled to a scalar field. In three
spacetime dimensions, the solution is exact for all $r$, whereas for
higher dimensions, the solution is an asymptotic one, for large $r$. 
The solutions have finite energy, although they do not possess an event 
horizon. The only non-zero charges are those associated with the 
timelike KVF. Furthermore, these charges contain information about the 
strength of the scalar field $k$. As we have discussed above, these 
results lead to interesting questions concerning the AdS/CFT conjecture. 
Among these are the issues of how a naked singularity on the gravity 
side translates to the field theory side, given that the associated 
energy is finite, and  the meaning of the energy degeneracy of the 
extremal solution and global AdS. Finally, it would be interesting to 
see if the energy degeneracies we have found can lead to the possibility 
of phase transitions analagous to the Hawking-Page transition \cite{hp} 
between Schwarzschild-AdS black holes and global AdS.

\smallskip
\noindent
{\it Acknowledgements} This work was supported  by the Natural Science
and Engineering Research Council of Canada.

\end{multicols}

\begin{thebibliography}{9}
\bibitem{maple} Equation (\ref{master}) may be integrated, for various 
dimensions and boundary conditions, using the following MAPLE code: 

 with(plots): k:=2: l:=1: d:=5:
  
 de := \{ p(r)\^{}2*(r*diff(p(r),r,r)/diff(p(r),r) 
 
 - ((d-1)*(d-2)*r\^{}2 + (d-3)*(d-4)*l\^{}2)/
 
 ((d-1)*r\^{}2+(d-3)*l\^{}2) ) - k\^{}2/3 \}:
  
 init := \{ p(10)=1, D(p)(10)=1 \}:
 
 psi := dsolve(de union init, p(r), type=numeric):
 
 odeplot(psi,[r,k/(p(r)*r)],0.1..10,numpoints=50);
\end{thebibliography}
\end{document}